\documentclass[twocolumn,twoside,slac_two]{revtex4}
\usepackage{graphicx}
\usepackage{fancyhdr}
\usepackage{amsmath}
\newcommand{\GeV}{{\,\rm GeV}}

\newcommand{\alpmz}{\alpha_s(m_Z)}
\newcommand{\pdff}{$(\partial F_{2} / \partial \ln  Q^{2})_x\,$ }
\newcommand{\gv}{GeV$^2\,$}
\def\etjet{E_T^{\rm jet}}
\def\set{d\sigma/d\etjet}
\newcommand{\alps}{\alpha_s}

\begin{document}

\title{$\alpha_s$ Determinations from Jets and Scaling Violations at HERA}

%

\author{T. Kluge}
\affiliation{DESY, Notkestr. 85, 22607 Hamburg, Germany}

\begin{abstract}
A review is given on recent $\alpha_s$ determinations from the H1 and ZEUS Collaborations.
These are based on measurements of jet cross sections, event shape variables, as well as on the observed scaling violation of the structure function $F_2$.
A HERA average on $\alpha_s(m_Z)$ is presented, in comparison with world mean values.   
\end{abstract}

\maketitle

\thispagestyle{fancy}

The strong coupling constant $\alpha_s$ is the single free parameter of QCD, the
knowledge of its value is essential when predicting
virtually every cross section for high energy collisions of elementary particles.
E.g.\ multi jet states pose the standard model background for various searches
for new physics at the LHC.
Thus it is essential to know  $\alpha_s$ to as high precision as possible.
The determination of the strong coupling is more difficult compared to other elementary forces
due to the confinement, i.e.\ one cannot observe the carriers of color charge, the partons, directly like electrons in QED.
Before the startup of LEP and HERA a precision of only around the $10\%$ level was reached \cite{Altarelli:1989ue}
\[\alpha_s(m_Z)=0.11\pm0.01 .\]
Much progress has been made since then, thanks to experimental data of higher precision, larger range in scale
and the inclusion of a variety of processes, such as
 $e^+e^-$ annihilation, hadron-hadron collisions,$ep$ scattering, heavy particle decays as well as 
advances in the theory.

HERA, the only accelerator for $e^\pm p$ collisions at high center-of-mass energies $\sqrt{s}=320\GeV$, is recognised
as today's precision tool for QCD investigations. 
With the H1 and ZEUS experiments two general purpose detectors explore the manifold aspects of QCD.
The role of HERA for determinations of the strong coupling constant lies in precision as well as in complementarity
to other environments.
In general two effects are exploited for this: scaling violations of structure functions (inclusive) and hadronic
final state characteristics (exclusive).

\begin{figure}
\includegraphics[width=55mm]{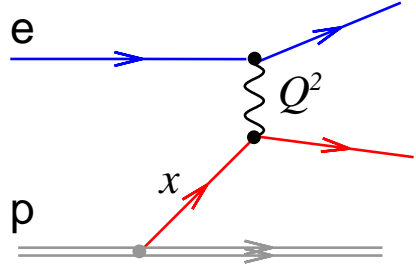}
\caption{The Born contribution to neutral current $ep$ scattering.}
\label{figI}
\end{figure}

\begin{figure}
\includegraphics[width=80mm]{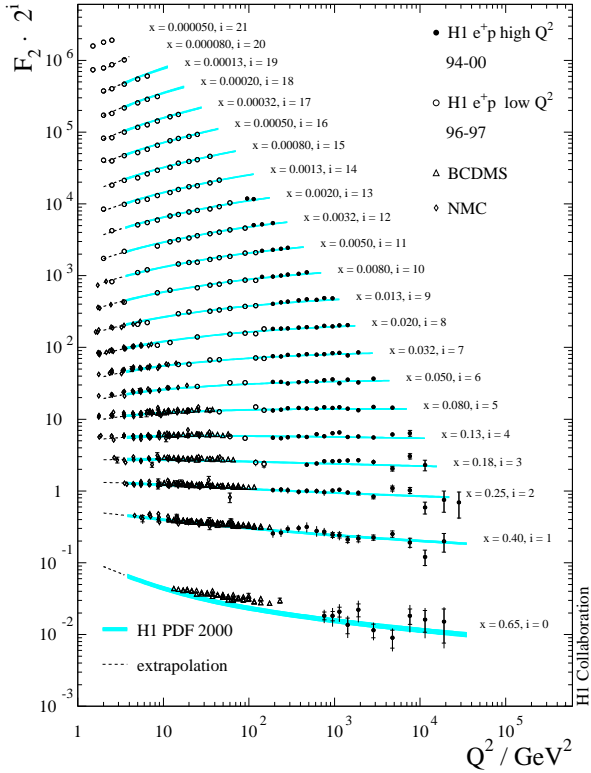}
\caption{The proton structure function $F_2$ measured by the H1 Collaboration compared with
  the corresponding Standard Model
  expectation determined from the H1 PDF $2000$ fit (error bands).
  Also shown are the $F_2$ data from BCDMS and NMC,
  which are not used in the fit.}
\label{fig1}
\end{figure}

\begin{figure}
\includegraphics[width=75mm]{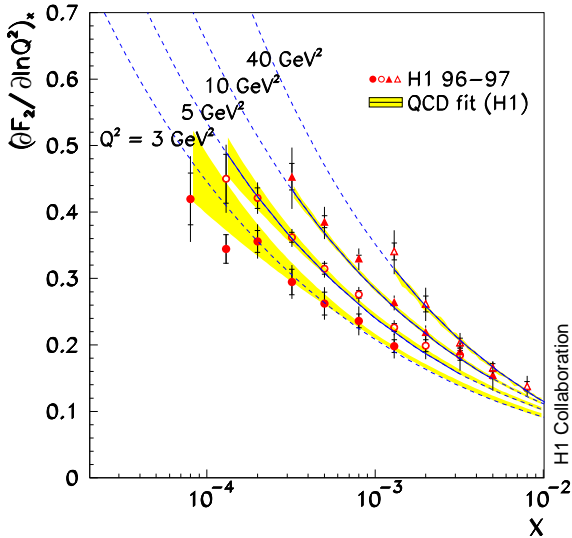}
\caption{The derivative \pdff  plotted
 as functions of $x$ for fixed $Q^2$,
for the H1 data  (symbols) and the QCD fit to
the H1 data, for $Q^2 \geq 3.5$~\gv (solid lines).
The error bands represent the model uncertainty of the
QCD analysis.}
\label{fig2}
\end{figure}

The cross section of the neutral current (NC) interaction
\mbox{$ep \rightarrow eX$} (Fig.~\ref{figI})
 is defined in terms
of three kinematic variables $Q^2$ (the photon virtuality), Bjorken $x$ and $y$, where $y$ quantifies
the inelasticity of the interaction.
 The kinematic variables are
related via $Q^2=sxy$, where $s$ is the $ep$ centre-of-mass energy squared.
Over most of the large kinematic domain at HERA the dominant contribution to
the cross section comes from pure photon exchange as expressed by
the structure function $F_2(x,Q^2)$, which is in the
quark parton model directly related to the sum of the quark and antiquark densities 
in the proton, i.e.\ $xF_2= \sum_i e_i^2(q_i(x,Q^2)+\bar q_i(x,Q^2))$.
In the quark parton model $F_2$ is independent of $Q^2$, a feature known as scaling, 
consequently the observation of scaling violations is evidence for QCD.
The accuracy and kinematic coverage of the HERA neutral and charged
current cross section data \cite{Adloff:2003uh,Breitweg:1999aa,Chekanov:2001qu}, provide for a clear demonstration of scaling violations,
e.g.\ shown in Fig.~\ref{fig1}.
The magnitude of the scaling violations depends on $\alps$ as well as on the 
gluon density in the proton.
The H1 Collaboration \cite{Adloff:2000qk} has determined  $\alpmz$ simultaneously with the 
gluon density 
in a QCD fit to the derivative of $F_2$ with respect to $Q^2$ (Fig.~\ref{fig2}),
using H1 and BCDMS proton data.
A value of the coupling constant
\begin{align*}
\alpha_s(m_Z)=0.1150\ \pm0.0017(\mathrm{exp.})&\ ^{+0.0009}_{-0.0005}(\mathrm{mod.})\\
 &\pm0.0050(\mathrm{th.})
\end{align*}
is obtained.
The theory error, estimated by varying the renormalisation scale and the factorisation scale  by a factor of four,
is found to be significantly larger than the experimental uncertainty.
 This uncertainty is expected to be reduced significantly in NNLO perturbation theory.
The value obtained for $\alpmz$ is nearly independent of
the chosen parameterisation for the parton densities.

\begin{figure}
\includegraphics[width=40mm]{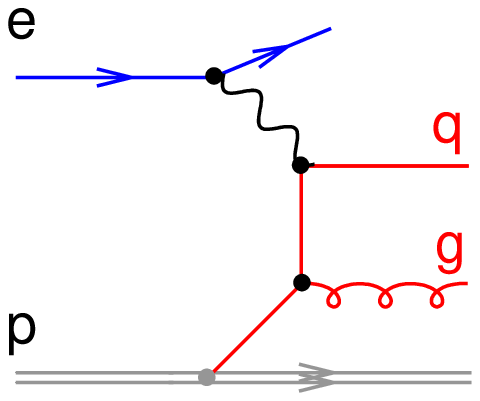}
\includegraphics[width=40mm]{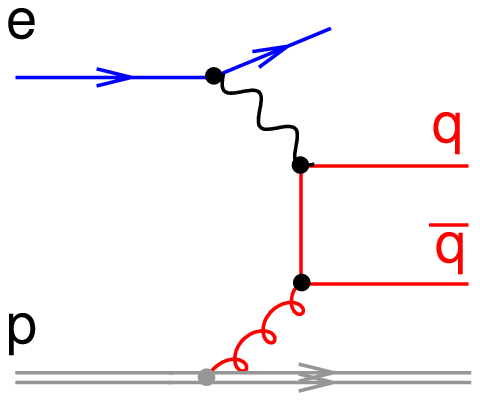}
\caption{The lowest order diagrams for two jet production in the Breit frame of reference.}
\label{figII}
\end{figure}

\begin{figure}
\includegraphics[width=65mm,angle=270]{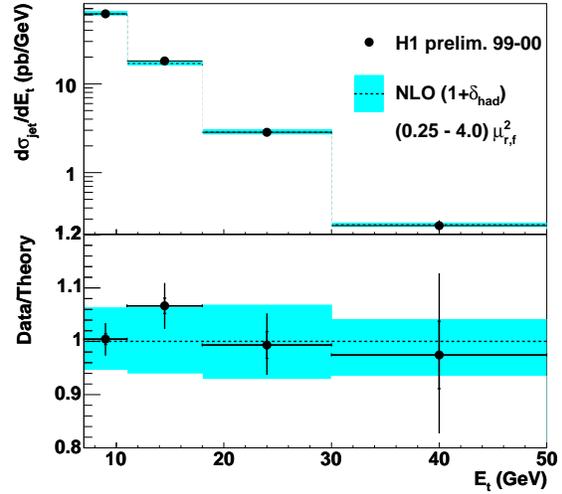}
\caption{Measured inclusive jet cross section $d\sigma/dE_t$ compared with NLOJET++ predictions corrected
for hadronisation effects. The bands show the theory uncertainty
by varying the renormalisation and factorisation scale by a factor of two.
The statistical uncertainty is shown as the inner error bar, while the total error is denoted by the outer error bar.}
\label{fig3}
\end{figure}

\begin{figure}
\includegraphics[width=55mm,angle=270]{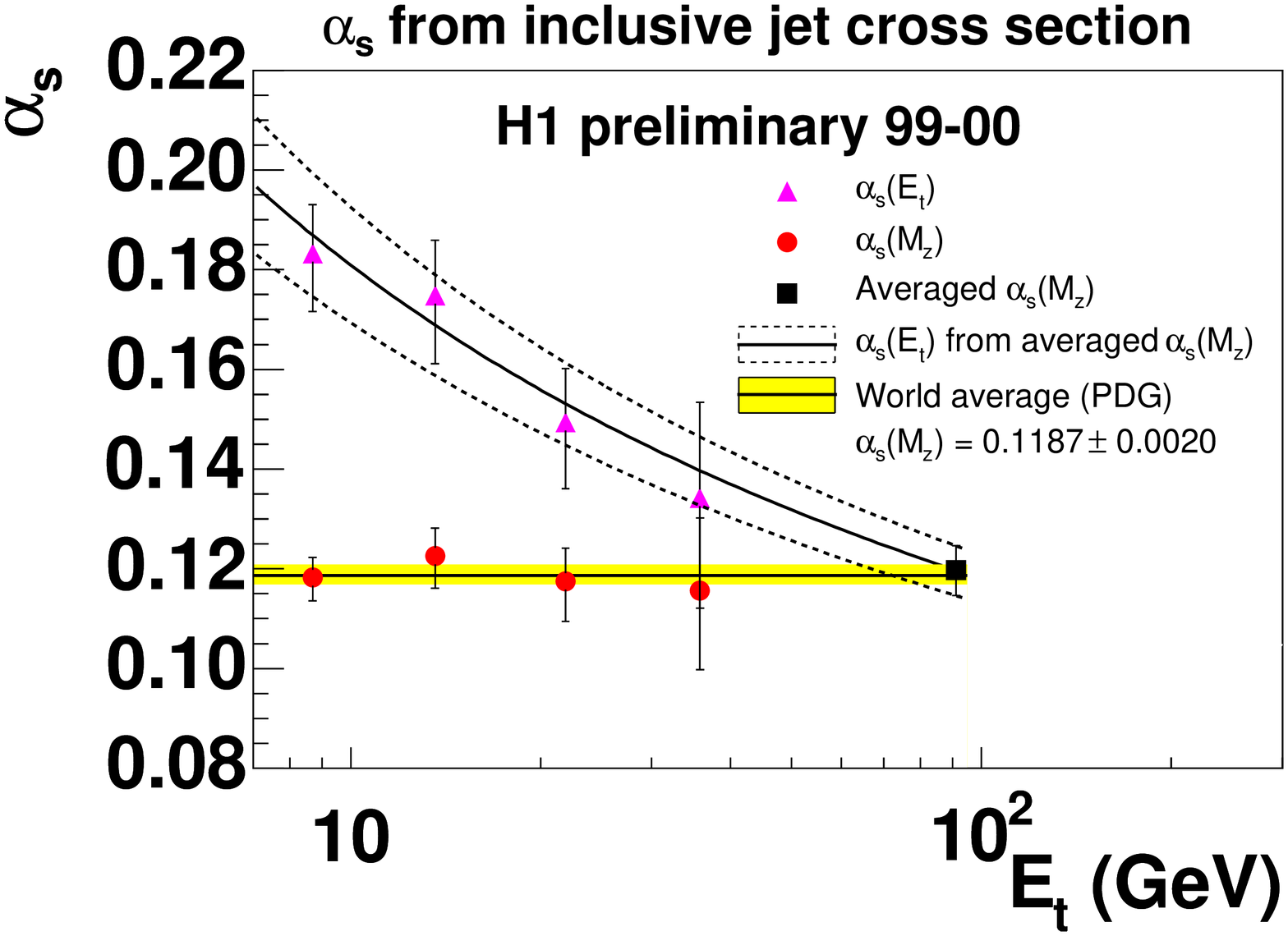}
\caption{
Results for  $\alpha_s(m_Z)$ and $\alpha_s(E_t)$,
where the $\alpha_s(E_t)$ values have been evolved from the
measured  $\alpha_s(m_Z)$ values for each bin in $E_t$ using the 
two loop solution of the renormalisation group equation.
The solid curve shows the result of evolving the averaged $\alpha_s(m_Z)$,
while the dashed curves indicate the respective error band.
The horizontal line with the shaded band corresponds to a world average.
}
\label{fig4}
\end{figure}

A more direct sensitivity to $\alps$ is achieved when including the hadronic final state of $ep$ scattering.
An example is the cross section for events where the final state contains more than one hadronic jet (not counting the 
remnant from proton dissociation), which
vanishes in absence of QCD effects (Fig.~\ref{figII}).
In deep-inelastic scattering (DIS, $Q^2\gtrsim 1\GeV^2$) the jets are usually reconstructed in the Breit frame of reference, where by definition
no recoil from the scattered beam electron occurs.
Jet cross sections are complementary compared to inclusive NC cross sections in that sense
that the analysis offers direct sensitivity to  $\alps$, on the other hand the measurement as well
 as the theoretical treatment is more involved, resulting to in general larger uncertainties of the data points.
The H1 and ZEUS Collaborations published results of inclusive jet and dijet cross sections \cite{Adloff:2000tq,Chekanov:2006xr}.
In a recent analysis the H1 Collaboration presented a measurement of the inclusive jet cross section in DIS \cite{H1:2005eps629} 
where the jets are defined by the $k_t$ algorithm in the Breit frame of reference.
The phase space spans high values of the photon virtuality $150<Q^2<5000\GeV^2$.
In  Fig.~\ref{fig3} the single differential cross section as function of the transverse energy $E_t$ of the jets
is shown.
A NLO calculation, corrected for hadronisation effects,
 gives a good description of the data over the full $E_t$ and $Q^2$ range. 
Hadronisation corrections are determined with Monte Carlo event generators,
they are found to be typically around $10\%$ from unity.
A fit of the strong coupling constant was performed by calculating the theory for several values of  $\alpmz$ 
and interpolating the cross section in between.
This interpolation allows for a mapping of the measured cross sections together with their
uncertainties onto a corresponding $\alpmz$ interval.
The result are shown in Fig.~\ref{fig4} as a function of the scale $E_t$.
The running of the strong coupling is clearly visible, and the values are found to be
consistent within errors to a world mean value.
Averaging the four individual values and taking into account correlated systematical errors
yields
\begin{align*}
\alpha_s(m_Z)=0.1197&\pm0.0016(\mathrm{exp.} )\\
&\ ^{+0.0046}_{-0.0048}(\mathrm{th.}).
\end{align*}
Again the error from theory is significantly larger than the experimental uncertainties.
The largest contribution to the latter comes from the hadronic energy scale of the main calorimeter.

\begin{figure}
\includegraphics[width=75mm]{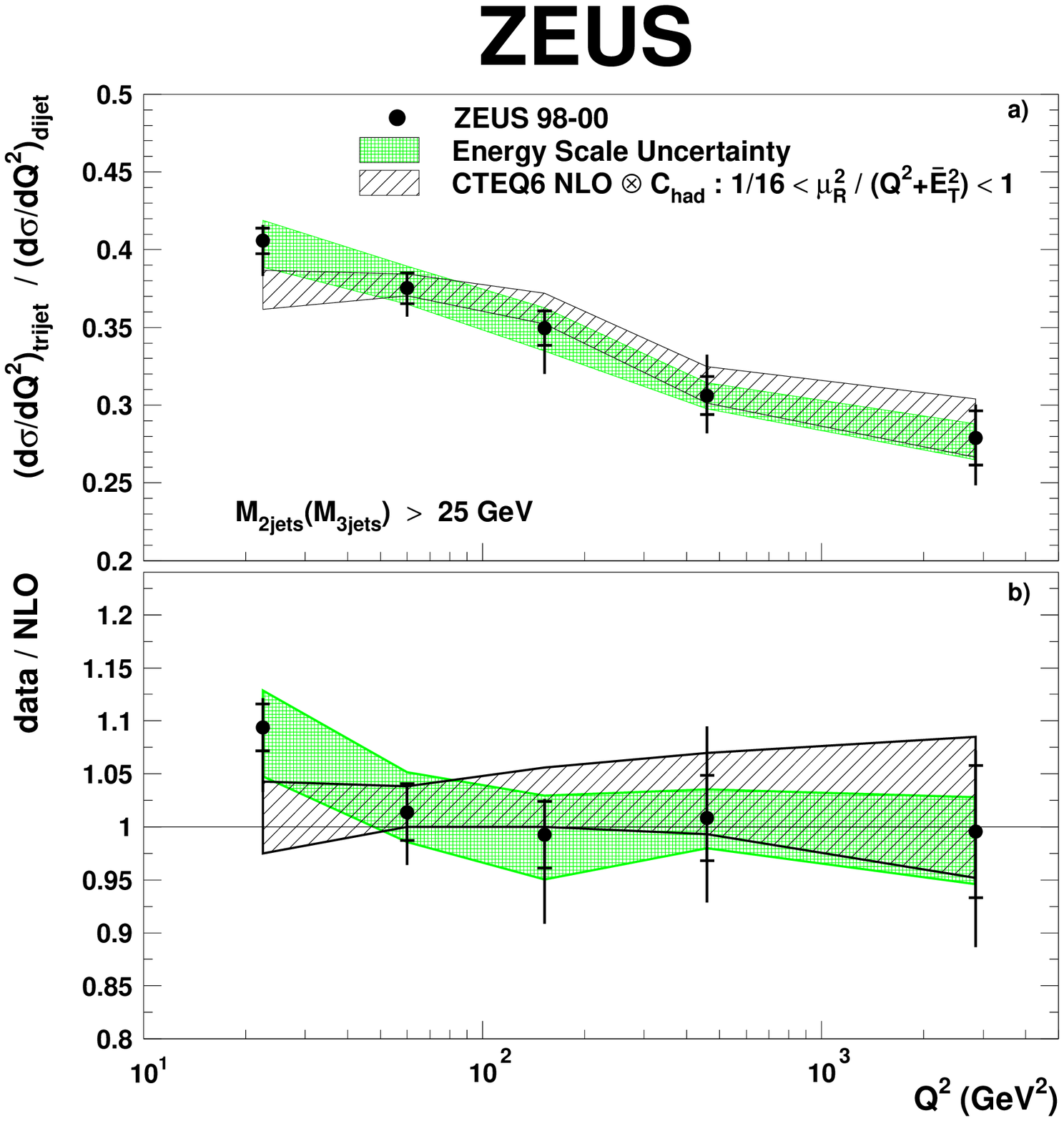}
\caption{ (a) The ratio of inclusive trijet to dijet cross sections as a
function of $Q^2$. The predictions of perturbative QCD in
next-to-leading order are compared to the data. (b) shows the ratio of the
data to the predictions. The inner error bars
represent the statistical uncertainties. The outer error bars
represent the quadratic sum of statistical and systematic uncertainties
not associated with the calorimeter energy scale. The shaded band
indicates the calorimeter energy scale uncertainty.}
\label{fig5}
\end{figure}

For inclusive jet cross sections every single jet is counted, alternatively the frequency of multi jet
events allows for another handle on the size of $\alps$.
One possibility is to study the cross section ratio of
three-jet and two-jet events, which is approximately proportional to $\alps$.
By construction a partial cancellation of errors, e.g.\ from the uncertainty of the gluon density in the proton, is expected.
There are results from H1 \cite{Kluge:2005im} as well as from ZEUS \cite{Chekanov:2005ve} available, which investigate the multi jet ratio ratio in DIS.
The  latter is shown as a function of the photon virtuality $Q^2$ in Fig.~\ref{fig5}.
The data are compared to a calculation at NLO precision, which has been corrected for hadronisation effects.
A good description within errors is obtained of both, the shape and the normalisation of the measured ratio.
The total experimental and theoretical
uncertainties are about $5\%$ and $7\%$, respectively.
These uncertainties are substantially reduced with respect to those of
the individual di- and trijet cross sections.
In particular, at lower $Q^2$ the theoretical uncertainties are reduced by as
much as a factor of four.
\begin{figure}
\includegraphics[width=75mm]{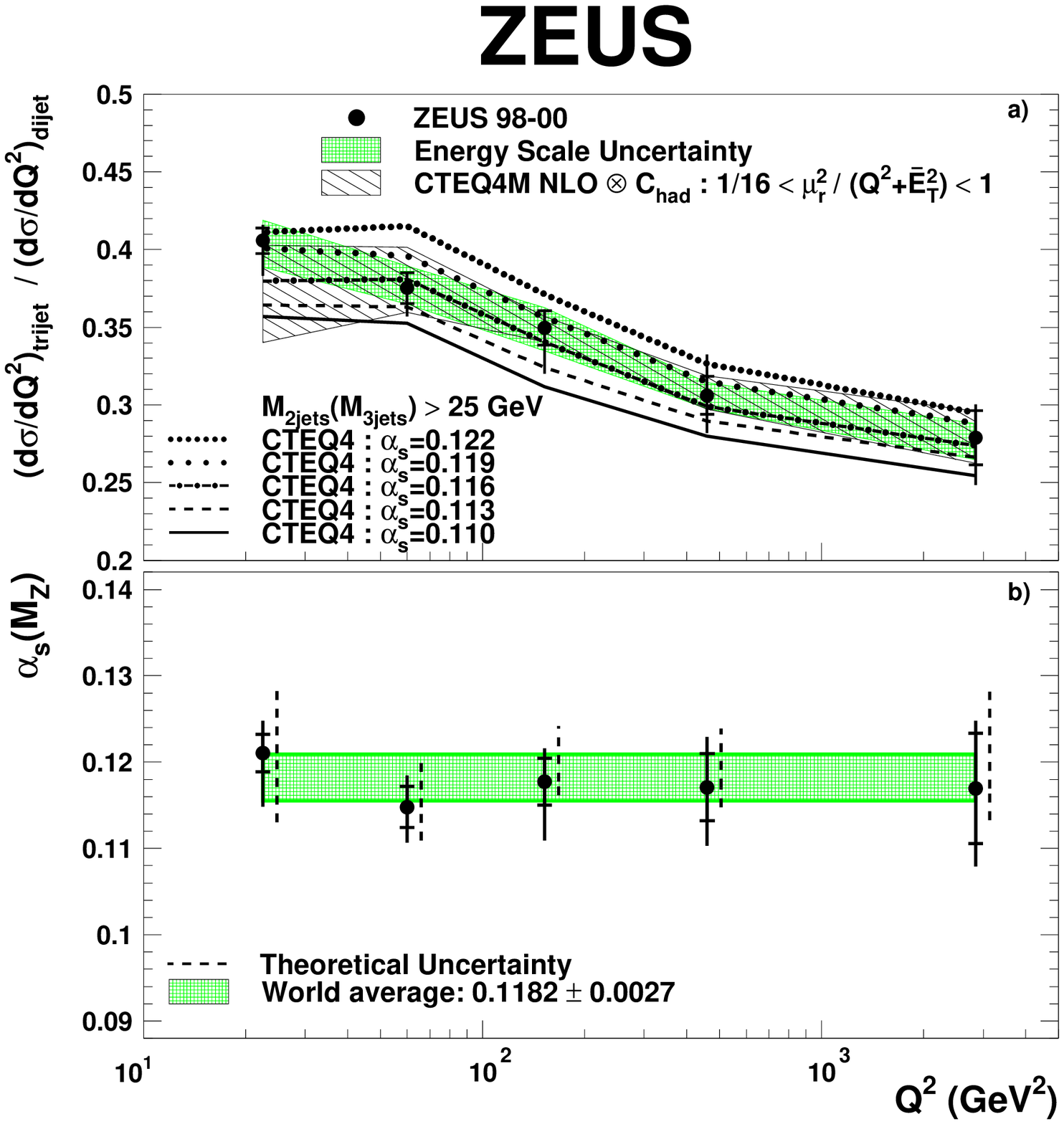}
\caption{(a) The ratio of inclusive trijet to dijet cross sections as a
function of $Q^2$. The predictions of perturbative QCD in
next-to-leading order using five sets of CTEQ4 PDF are compared to the data.
(b) shows the $\alpha_s(M_Z)$ values determined from the
ratio of inclusive trijet to dijet cross
sections in different regions of $Q^2$.
The shaded band indicates a world average value of $\alpha_s(M_Z)$.
The dashed error bars display the theoretical uncertainties.}
\label{fig6}
\end{figure}
In the upper diagram of Fig.~\ref{fig6} the sensitivity to the strong coupling is highlighted.
The theory prediction is shown for different values of $\alpmz$.
A method similar to what was used for the inclusive jets yields values of $\alpmz$ as shown in the lower diagram.
The mean value
\begin{align*}
\alpha_s(m_Z)=0.1179\ \pm0.0013(\mathrm{stat.} )&\ ^{+0.0028}_{-0.0046}(\mathrm{exp.})\\
&\ ^{+0.0064}_{-0.0046}(\mathrm{th.}) 
\end{align*}
is found to be compatible with what was extracted from the inclusive jets, where the latter exhibit a somewhat smaller uncertainty.
On the other hand, the ratio of multi jet cross sections allows for an extension of the phasespace to 
lower $Q^2$ ($10\GeV^2$ compared to $150\GeV^2$), due to cancellations
of the theoretical uncertainties. 

\begin{figure}
\includegraphics[width=75mm]{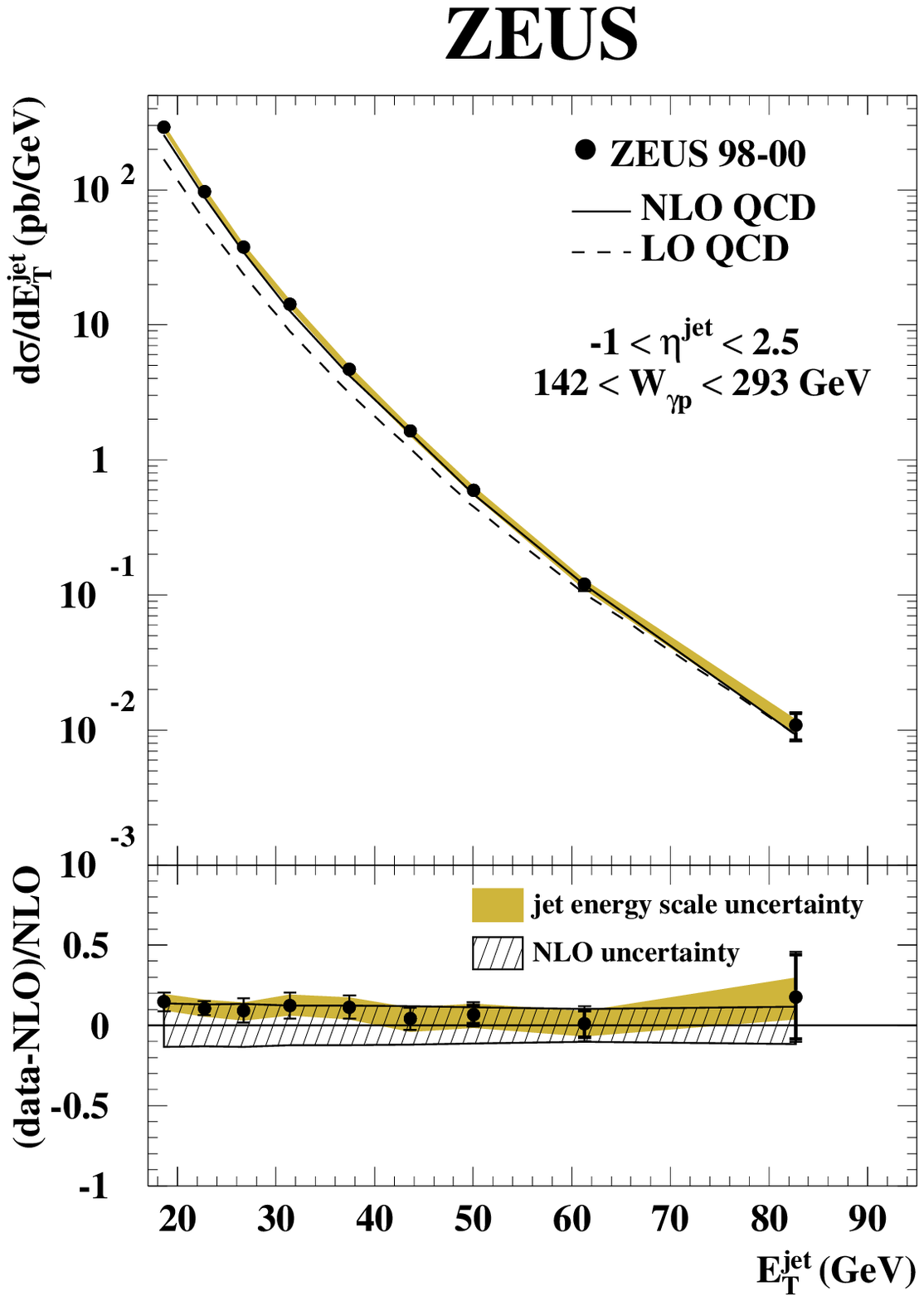}
\caption{The inclusive jet cross section $\set$ in photoproduction.
The uncertainty associated with the absolute
energy scale of the jets is shown as a shaded band.
The LO (dashed line) and NLO (solid line) QCD parton-level
calculations corrected for hadronisation effects
are also shown. The lower diagram shows the fractional
difference between the measured $\set$ and the NLO QCD calculation;
the hatched band shows the uncertainty of the calculation.}
\label{fig7}
\end{figure}

In the regime of photo production at even lower $Q^2<1\GeV^2$ an additional
resolved component to the cross section needs to be taken into account.
In that case the photon acts like a hadron as a source of partons, which in the theory 
is reflected by the introduction of a second set of parton density functions.
The large cross section for photo production allows to extend the measurement to higher jet transverse energies.
An analysis by the ZEUS Collaboration \cite{Chekanov:2002ru} spans $17<E_t<95\GeV$,
shown in Fig.~\ref{fig7}.
The measured $\set$ falls by over five orders of magnitude in this $\etjet$ range.
Fixed-order QCD calculations are compared to the data, where the LO QCD calculation underestimates the
measured cross section by about $50\%$ for
$\etjet<45$ GeV.
The calculation that includes NLO corrections gives a
good description of the data within the experimental and theoretical
uncertainties over the complete $\etjet$ range studied.

A value of $\alpmz$ is obtained by combining all the $\etjet$ regions
\begin{align*}
\alpha_s(m_Z)=0.1224\ \pm0.0001(\mathrm{stat.} )&\ ^{+0.0022}_{-0.0019}(\mathrm{exp.})\\
&\ ^{+0.0054}_{-0.0042}(\mathrm{th.}) ,
\end{align*}
which is consistent with the determinations from jet production in NC DIS.
The largest contribution to the experimental uncertainty comes from the jet energy scale and amounts to $\pm 1.5\%$
on $\alpmz$.
Among the contributions to the theoretical uncertainty on the strong coupling, the largest one is due to 
terms beyond NLO, which is estimated as ${}^{+4.2}_{-3.3}\%$ by varying the renormalisation and factorisation scales by a factor of four.
Uncertainties connected to the proton and photon PDFs and to the hadronisation were significantly smaller.

\begin{figure}
\includegraphics[width=75mm]{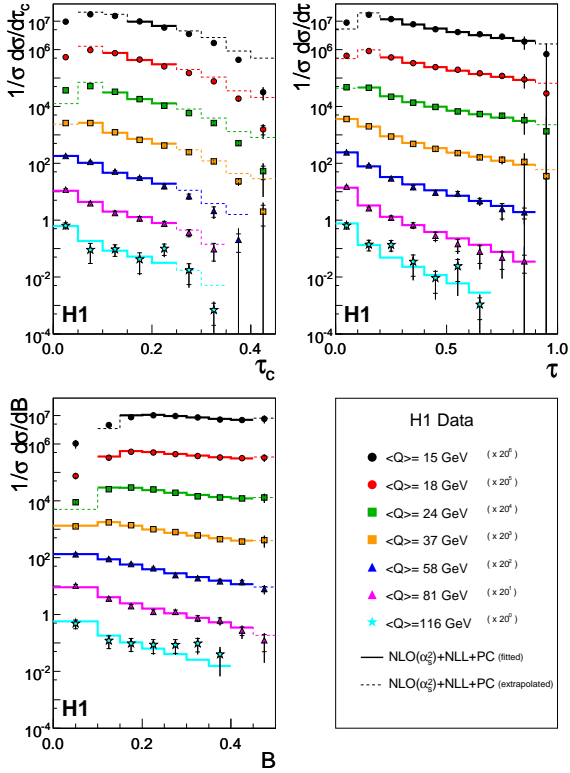}
\caption{ Normalised event shape distributions corrected to the hadron level
    for the $\tau_C$, $\tau$ and $B$ variables.
    The measurements are compared with fits based on a NLO QCD calculation including 
    resummation (NLL) and supplemented by power corrections (PC).
    The fit results are shown as solid lines and are extended as
    dashed lines to those data points which are not included in the QCD fit.}
\label{fig8}
\end{figure}

\begin{figure}
\includegraphics[width=75mm]{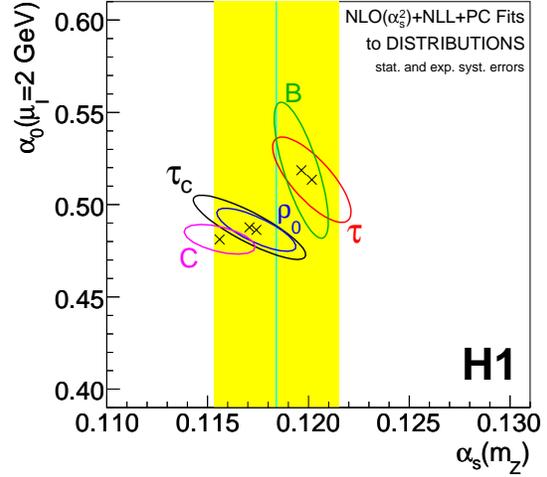}
\caption{Fit results to the differential distributions of
    $\tau$, $B$, $\rho_0$, $\tau_C$ and the $C$-parameter 
    in the $(\alpha_s,\alpha_0)$ plane.
    The $ 1\sigma$ contours
    correspond to $\chi^2 = \chi^2_{\rm min}+1$,
    including statistical and experimental systematic uncertainties.
    The value of $\alpha_s$ (vertical line) and its uncertainty (shaded band) are taken from \cite{Bethke:2004uy}.}
\label{fig9}
\end{figure}

\begin{figure}
\includegraphics[width=75mm]{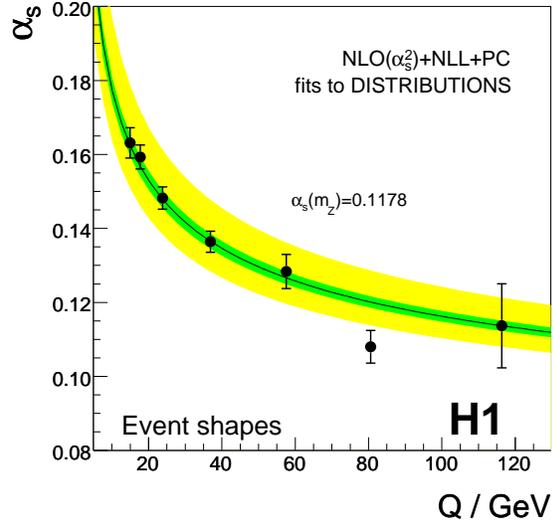}
\caption{ The strong coupling $\alps$ as a function of the scale $Q$ from 
    an average of the results obtained by fitting the
differential event shape distributions.
The errors represent the total experimental uncertainties.
A value of $\alps(m_Z)$ is indicated in the plot, determined from a fit to the  $\alps(Q)$ results using the
QCD renormalisation group equation.
The fit curve is shown as the full line.
The inner (outer) shaded band represents the uncertainty of the fitted 
$\alps(Q)$ from experimental errors (the renormalisation scale variation).}
\label{fig10}
\end{figure}

Event shape variables play an important role in QCD studies of the hadronic final state.
They are a class of topological observables which are defined as positive real numbers
 calculated from the four vectors of all hadronic final state particles.
Both, the H1 and ZEUS Collaborations \cite{Aktas:2005tz,Chekanov:2006hv} published recently new investigations
of event shape variables, motivated by new developments from the theory side.
There are several differences compared to jet cross sections: larger statistics due to the
semi-inclusive nature of event shapes, reduced experimental systematic uncertainties from
hadronic energy scales and larger hadronisation effects.
The hadronisation needs a special treatment, in recent DIS studies by analytic power corrections.
In general event shapes variables are defined to be equal to zero without QCD effects and are of order one for multi jet configurations, 
which can occur due to multiple gluon radiation.
For differential distributions, fixed order calculations alone proved to be not sufficient, 
 hence soft gluon resummations are supplementing the NLO predictions.
The H1 Collaboration \cite{Aktas:2005tz} published distributions of five event shape variables, shown in Fig.~\ref{fig8}.
Towards higher scales $Q$ the distributions are more peaked near zero (the value for the quark parton model), 
which gives evidence for asymptotic freedom of QCD.
This distributions are compared to calculations based on fixed order and matched resummed parts.  
To take into account hadronisation Dokshitzer/Webber power corrections (PC) have been used, which 
depends on the parameter $\alpha_0$ representing an effective strong coupling constant in the infrared regime.  
An overall good description is obtained for part of the phase space (higher $Q$ and moderate event shape values),
 where the theory is expected to be valid.
Simultaneous fits of $\alpmz$ and the power correction parameter $\alpha_0$ are shown in Fig.~\ref{fig9}.
An average value of
\begin{align*}
\alpha_s(m_Z)=0.1198&\ \pm0.0013(\mathrm{exp.} )\\
&\ ^{+0.0056}_{-0.0043}(\mathrm{th.})
\end{align*}
is obtained,
which is consistent with the results from jet and inclusive DIS cross sections.
The fit was also performed separately for all scales covered by the data, see Fig.~\ref{fig10}, where the 
 asymptotic freedom of QCD is clearly demonstrated.
Due to the more inclusive definition compared to jets, a larger range in scale is accessible for the event shape analysis.

\begin{figure}
\includegraphics[width=75mm]{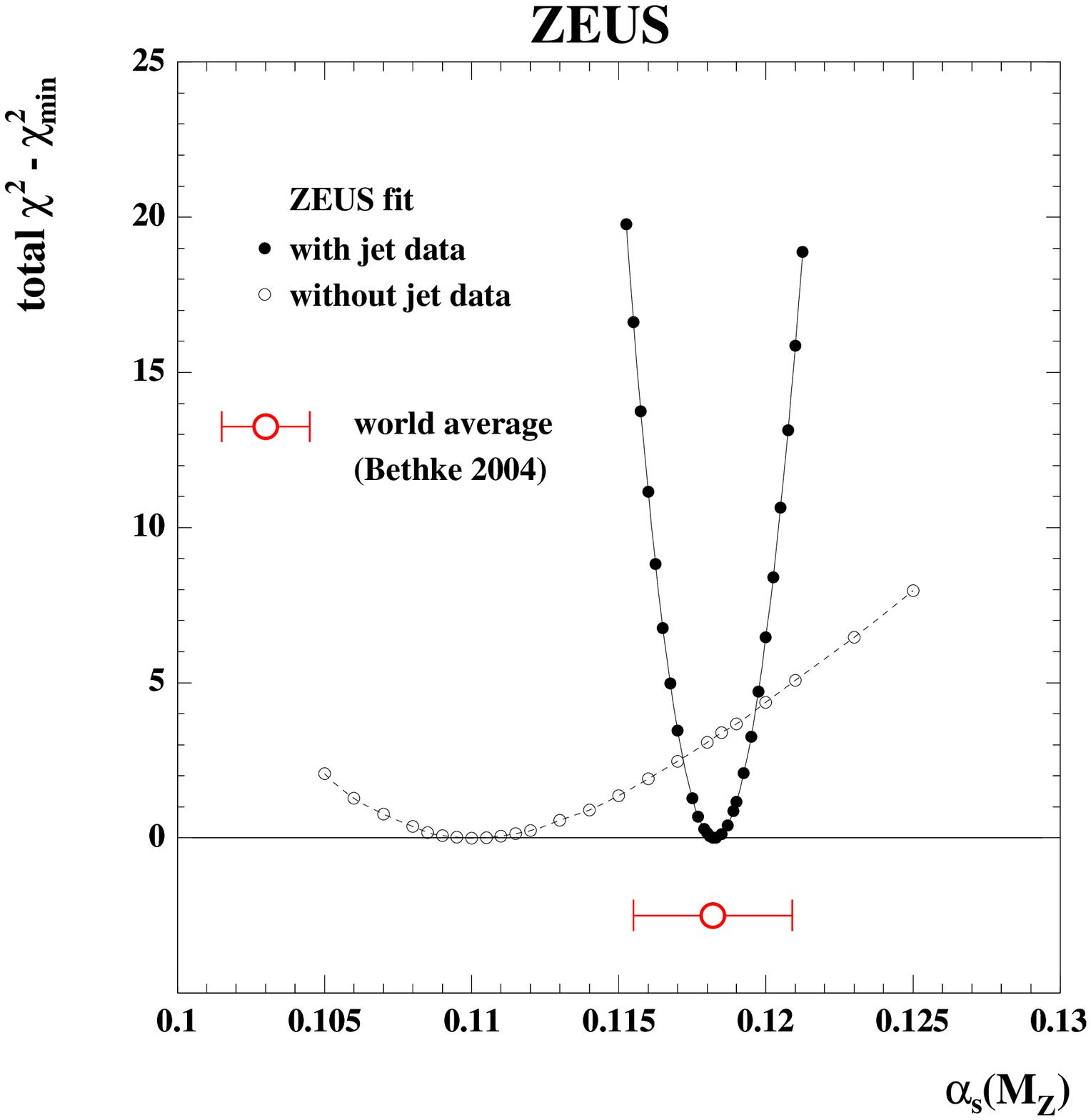}
\caption{The $\chi^2$ profile as a function of $\alpha_s(m_Z)$ for the
ZEUS-JETS-$\alpha_s$ fit (black dots) and for a similar fit not including the jet data
(clear dots). The ordinate is given in terms of the difference between the
total $\chi^2$ and the minimum $\chi^2$, for each fit.}
\label{fig11}
\end{figure}

\begin{figure}
\includegraphics[width=75mm]{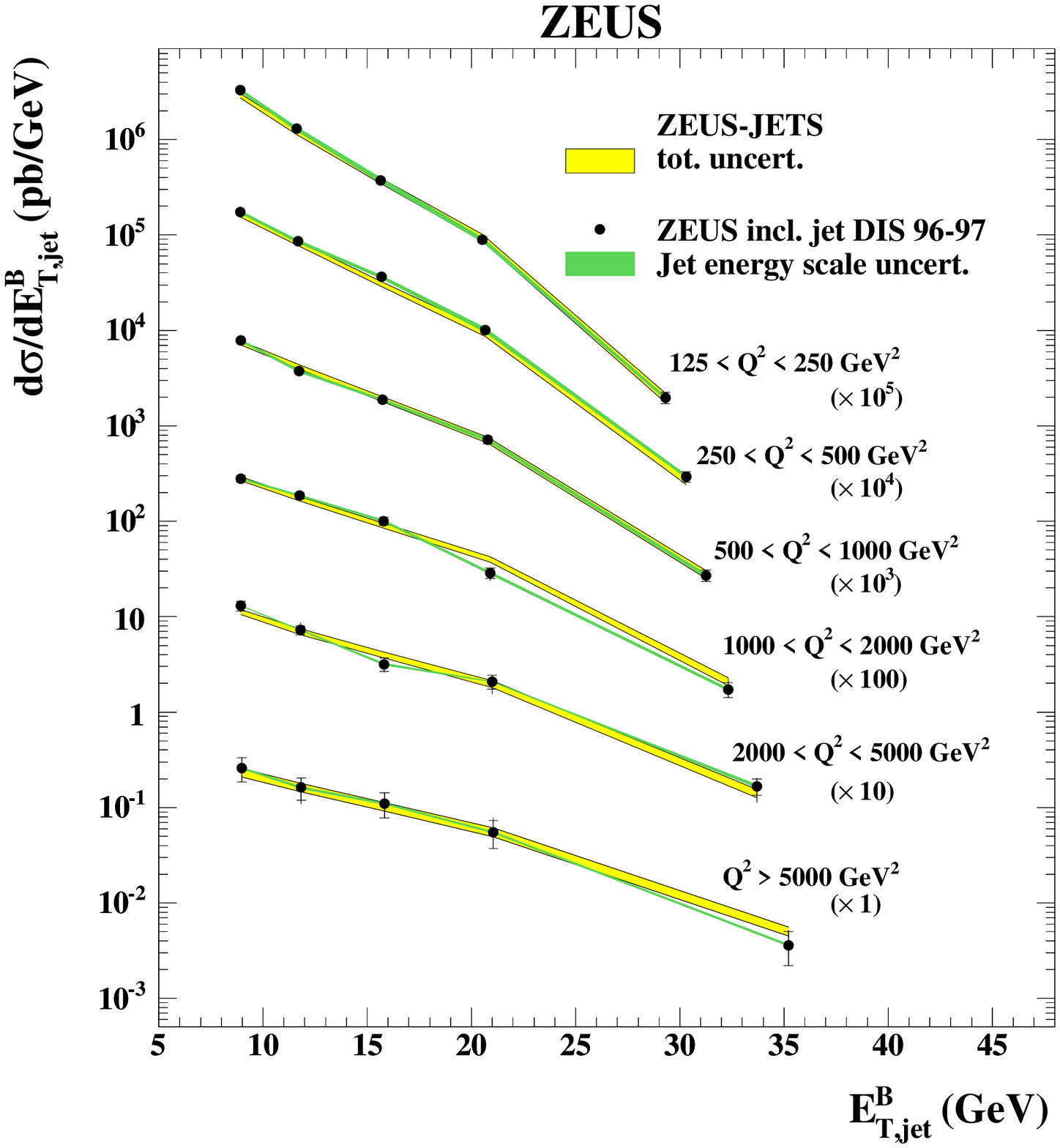}
\caption{ZEUS-JETS fit compared to ZEUS DIS jet data. Each cross section
has been multiplied by the scale factor in brackets to aid visibility.}
\label{fig12}
\end{figure}

Since inclusive DIS and jet analyses offer different sensitivity to the PDFs of the proton and $\alps$, it is desirable 
to have a combined QCD analysis based on both classes of observables, simultaneously fitting the PDFs and the strong coupling. 
The addition of jet data allows an extraction of
$\alpha_s(m_Z)$ that is less correlated to the shape of the gluon density in the proton.
Such an investigation was presented by the ZEUS Collaboration \cite{Chekanov:2005nn}.
A clear improvement on the uncertainty of $\alpmz$ by adding jet data to the inclusive fit is evident from Fig.~\ref{fig11}.
The included jet data comprise DIS as well as photo production data, an example is shown in Fig.~\ref{fig12}.
The result of the combined fit for the strong coupling yields
\begin{align*}
\alpha_s(m_Z)=0.1183\ \pm0.0028(\mathrm{exp.} )&\pm0.0008(\mathrm{mod.} )\\
&\pm0.005(\mathrm{th.} ),
\end{align*}
where in contrast to the determinations given before the theory uncertainty was estimated by varying
the scale by a factor of two (instead of four).

\begin{figure}
\includegraphics[width=85mm]{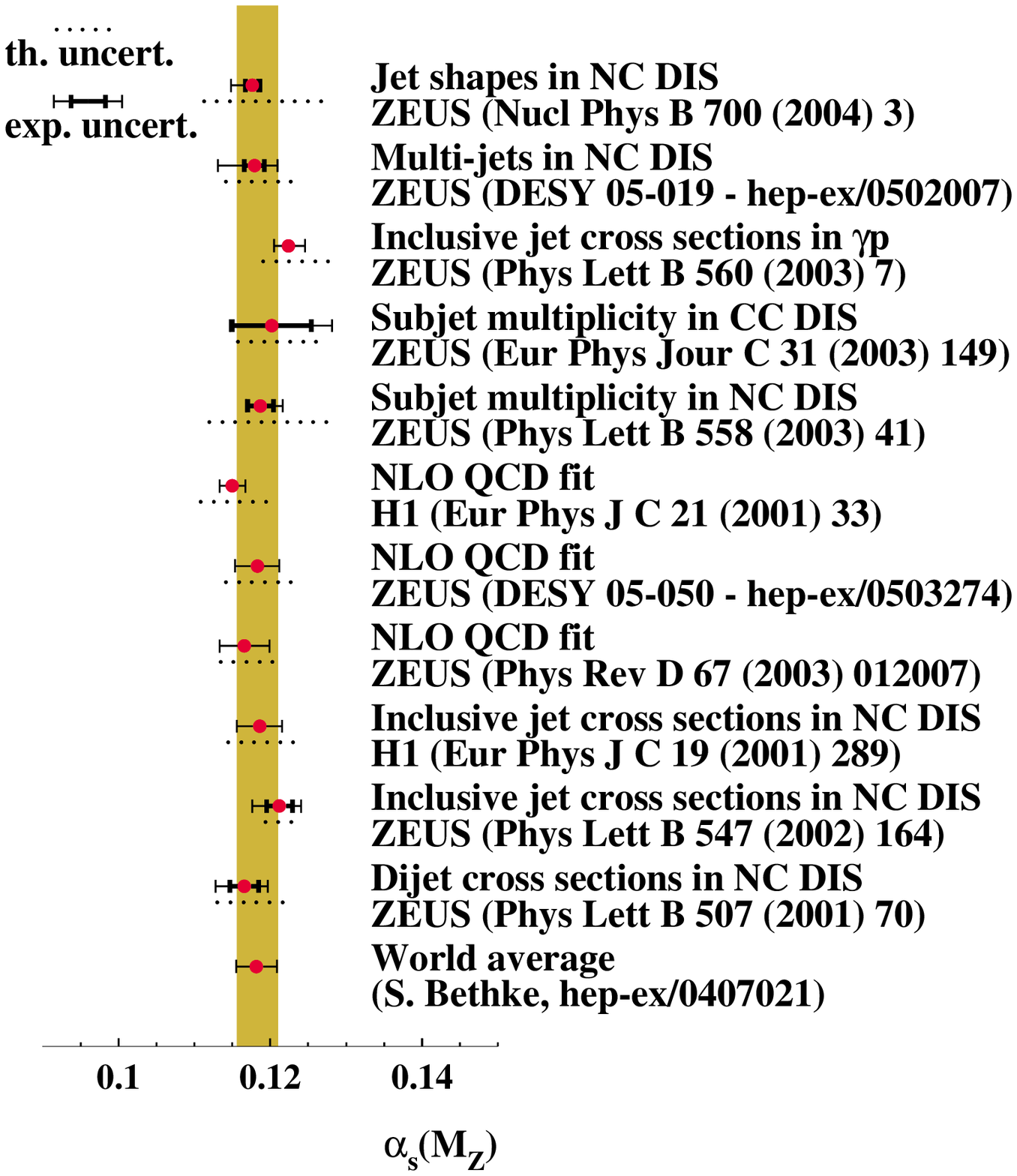}
\caption{ Summary of $\alpha_s(m_Z)$ determinations at HERA compared with
a world average.}
\label{fig13}
\end{figure}

From what was presented in this article so far it is evident that
there is a great diversity of $\alpmz$ determinations from the H1 and ZEUS Collaborations.
An HERA average from selected values was constructed in \cite{Glasman:2005ik}.
The individual results which entered the average are listed in Fig.~\ref{fig13}.
All of these are consistent with each other and the theory uncertainty always dominates
the total error.
The average was build by taking into account the known
correlations in each experiment and assuming conservatively that the
theoretical uncertainties arising from the terms beyond NLO are fully
correlated.
The HERA average
\begin{align*}
\alpha_s(m_Z)=0.1186\ &\pm0.0011(\mathrm{exp.})\\
&\pm 0.0050(\mathrm{th.})
\end{align*}
is compatible with world means (e.g. from \cite{Bethke:2004uy}) and offers competitive precision.

\begin{figure}
\includegraphics[width=65mm]{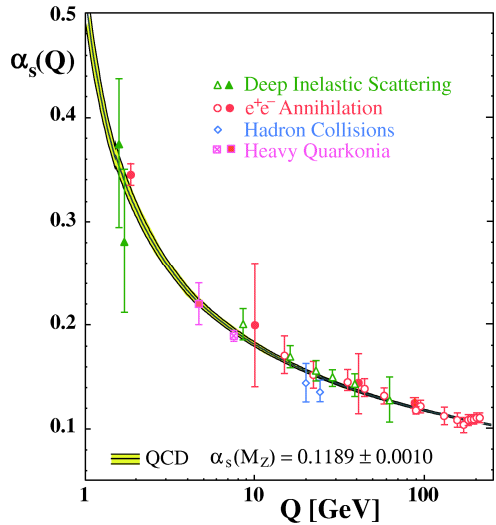}
\caption{Summary of measurements of $\alps (Q)$ as a function of the
respective energy scale $Q$ \cite{Bethke:2006ac}.
Open symbols indicate (resummed) NLO, and
filled symbols NNLO QCD calculations used in the respective
analysis.
The curves are the QCD predictions for the combined world
average value of $\alpmz$, in 4-loop approximation and using 3-loop
threshold matching at the heavy quark pole masses
$M_c = 1.5$~GeV and $M_b = 4.7$~GeV.}
\label{fig14}
\end{figure}

This HERA average was included in a more recent world average
\cite{Bethke:2006ac}, 
\[
\alpha_s(m_Z)=0.1189\pm0.0010,
\]
where for the first time jet data from HERA were considered.
The individual contributions as a function of their scale are 
shown in Fig.~\ref{fig14}
The total error on this average is quoted to be  below $1\%$, a large improvement compared to previous world means,
which could be reached by including contributions from tau decays and lattice QCD.

The uncertainties of $\alpmz$ determinations from HERA data alone are somewhat larger, still compared to $e^+e^-$ annihilation and hadron collisions
they are competitive.
Moreover, the observed compatibility of the results from HERA and other processes is essential to support the universality of QCD.

\bigskip 
\bibliography{tkbib}





\end{document}